\pdfoutput=1
\documentclass[sigconf]{acmart}
\usepackage[english]{babel}
\usepackage{blindtext}
\usepackage{tabularx}
\usepackage{mathtools}
\usepackage{amsmath}
\usepackage{epsfig,endnotes}
\usepackage{url}
\usepackage{multirow}
\usepackage{placeins}
\usepackage{color}
\usepackage{balance}
\usepackage{tikz}
\usepackage{algorithm}
\usepackage{algpseudocode}
\usepackage{hyperref}
\usepackage{booktabs}
\usepackage{graphicx}
\usepackage{rotating}
\usepackage{verbatim}
\usepackage{xcolor}
\usepackage{caption}
\usepackage{subcaption}
\usepackage[english]{babel}
\usepackage[utf8]{inputenc}
\usepackage{array, makecell}
\usepackage{xspace}
\usepackage{tablefootnote}
\usepackage{colortbl}

\renewcommand\footnotetextcopyrightpermission[1]{}
\newcommand\blfootnote[1]{%
  \begingroup
  \renewcommand\thefootnote{}\footnote{#1}%
  \addtocounter{footnote}{-1}%
  \endgroup
}

\newcommand{\para}[1]{{\vspace{1pt} \bf \noindent #1 \hspace{1pt}}}

\newcommand{\eg}{{\em e.g.,\ }}
\newcommand{\ie}{{\em i.e.,\ }}

\definecolor{lcyan}{RGB}{1,0,0}

\newenvironment{packed_itemize}{
\begin{list}{\labelitemi}{\leftmargin=1.em}
  \setlength{\itemsep}{2pt}
  \setlength{\parskip}{0pt}
  \setlength{\parsep}{0pt}
  \setlength{\headsep}{0pt}
  \setlength{\topskip}{0pt}
  \setlength{\topmargin}{0pt}
  \setlength{\topsep}{0pt}
  \setlength{\partopsep}{0pt}
}{\end{list}}

\begin{document}

\title{Deepfake Videos in the Wild: Analysis and Detection}




\author{Jiameng Pu}
\authornotemark[2]
\affiliation{
\institution{Virginia Tech}
\country{}}
\email{jmpu@vt.edu}

\author{Neal Mangaokar}
\authornotemark[2]
\affiliation{\institution{University of Michigan}
\country{}}
\email{nealmgkr@umich.edu}

\author{Lauren Kelly}
\affiliation{\institution{Virginia Tech}
\country{}}
\email{bmkelly@vt.edu}

\author{Parantapa Bhattacharya}
\affiliation{\institution{University of Virginia}
\country{}}
\email{parantapa@virginia.edu }

\author{Kavya Sundaram}
\affiliation{\institution{Virginia Tech}
\country{}}
\email{smkavya@vt.edu}

\author{Mobin Javed}
\affiliation{\institution{LUMS Pakistan}
\country{}}
\email{mobin.javed@lums.edu.pk}

\author{Bolun Wang}
\affiliation{\institution{Facebook}
\country{}}
\email{bolunwang@fb.com}

\author{Bimal Viswanath}
\affiliation{\institution{Virginia Tech}
\country{}}
\email{vbimal@cs.vt.edu}



\newcommand{\parait}[1]{{\hspace{-7.5pt} \vspace{1pt} \it \noindent #1 \hspace{6pt}}}

\newcommand{\wilddataset}[1]{{DF-W}}
\newcommand{\researchdataset}[1]{{DF-R}}

\renewcommand{\shortauthors}{Pu and Mangaokar, et al.}

\begin{abstract}
AI-manipulated videos, commonly known as deepfakes, are an emerging problem. Recently, researchers in academia and industry have contributed several (self-created) benchmark deepfake datasets, and deepfake detection algorithms. However, little effort has gone towards understanding deepfake videos in the wild, leading to a limited understanding of the real-world applicability of research contributions in this space. Even if detection schemes are shown to perform well on existing datasets, it is unclear how well the methods generalize to real-world deepfakes. To bridge this gap in knowledge, we make the following contributions: \textit{First}, we collect and present the largest dataset of deepfake videos in the wild, containing 1,869 videos from YouTube and Bilibili, and extract over 4.8M frames of content. \textit{Second}, we present a comprehensive analysis of the growth patterns, popularity, creators, manipulation strategies, and production methods of deepfake content in the real-world. \textit{Third}, we systematically evaluate existing defenses using our new dataset, and observe that they are not ready for deployment in the real-world. \textit{Fourth}, we explore the potential for transfer learning schemes and competition-winning techniques to improve defenses.
\end{abstract}


\maketitle
\pagestyle{plain}

\section{Introduction}
\label{sec:intro}

Advances in deep neural networks (DNNs) have enabled new ways of manipulating video content. This has fueled the rise of \textit{deepfakes}, or videos where the face of a person is swapped in by another face, using DNN-based methods. Popular DNN methods for creating deepfakes include  Autoencoders~\cite{rumelhart1985learning}, Generative Adversarial Networks (GANs)~\cite{goodfellow2014generative}, and Variational Autoencoders (VAEs)~\cite{kingma2014vae}. Such capabilities to produce convincing deepfakes raises serious ethical issues, because they can be misused in many ways, \eg to show a person at a place they never went, make a person perform actions they never did, or say things they never said. Such fake videos can help spread fake news, manipulate elections, or incite hatred towards minorities~\cite{agarwal2019protecting}.

Given the potential for misuse of this technology, researchers have proposed a variety of deepfake video detection schemes~\cite{guera2018deepfake, UADFV, afchar2018mesonet, nguyen2019use, zhou2017two, nguyen2019multi, yang2019exposing, agarwal2019protecting}, and also released new deepfake datasets~\cite{UADFV, deepfaketimit, rossler2019faceforensics++, googledfd, li2019celeb, dolhansky2020deepfake, jiang2020deeperforensics} to evaluate the proposed detection schemes. In addition, governments are also realizing the seriousness of this issue. In the USA, the first federal legislation on deepfakes was signed into law in December 2019~\cite{federaldeepfakelaw}, with the government encouraging research on deepfake detection technologies.
However, most existing research efforts from academia and industry have been conducted with limited or no knowledge of actual \textit{deepfake videos in the wild}. Therefore, we have a limited understanding of the real-world applicability of existing research contributions in this space. A number of questions can be raised: (1) \textit{How are deepfake videos created in the wild? (2) Are deepfakes in the wild different from deepfake videos produced by the research community? (3) Are deepfakes increasingly appearing in the wild? Are they being viewed by large populations? (4) Can existing deepfake detection schemes (primarily tested on deepfakes produced by the research community) accurately detect deepfake videos in the wild?}

\blfootnote{$\dagger$ indicates equal contribution.}
This work aims to answer these questions by conducting a large-scale measurement study of deepfake videos in the wild or deepfake videos produced and shared by the Internet community (\ie not researchers). To the best of our knowledge, this is the largest measurement study to date. Our contributions include the following: 
\begin{packed_itemize}
\item \textit{We introduce a new deepfake dataset called \wilddataset{}, comprised of deepfake videos created and shared by the Internet community.} We prepare this dataset by scanning a variety of sources on the Web, including YouTube, Bilibili~\cite{bibilibilink}, and Reddit.com. Our dataset includes a total of $1,869$ videos from YouTube and Bilibili, comprising of over 48 hours of video, covering a wide range of video resolutions. To the best of our knowledge, \wilddataset{} is the \textit{largest} collection of deepfake videos in the wild.

\item \textit{We present a comprehensive analysis of the videos in \wilddataset{}.} We examine the differences in content between deepfake videos in \wilddataset{}, and datasets released by the research community~\cite{UADFV, deepfaketimit, rossler2019faceforensics++, googledfd, li2019celeb, dolhansky2020deepfake}. We observe that \wilddataset{} videos tend to be more sophisticated, and include several variations of deepfake content, thus raising new challenges for detection schemes. We find that many \wilddataset{} videos are created using generation methods different from those used by the research community, which potentially results in a data distribution gap between existing deepfake datasets and \wilddataset{} deepfakes.
We also analyze the growth, and popularity of deepfake videos in the wild, and investigate the content creators involved in the process.

\item \textit{We systematically evaluate the performance of state-of-the-art deepfake detection schemes on videos in \wilddataset{}.} We evaluate 7 deepfake detection schemes, including 5 supervised and 2 unsupervised schemes. We find that all detection schemes perform poorly on \wilddataset{} videos, with the best approach (CapsuleForensics, a supervised approach) having an F1 score of only 77\% in catching deepfakes.
This means that these existing detection schemes are not ready for real-world deployment. Poor performance can be attributed to distributional differences between real-world deepfake videos, and those used to train existing detection schemes.
Failure cases can also be partially attributed to racial bias, a well known problem with DNN-based facial analysis~\cite{robinson2020face}. We also attempt to interpret the classification decisions using a state-of-the-art model interpretation scheme, called Integrated Gradients~\cite{sundararajan2017axiomatic}. We leverage this tool to infer features that can be used to either improve detection schemes, or create more evasive deepfakes. 

\item \textit{We explore approaches to improve detection performance.}
Finally, to improve detection performance on \wilddataset{}, we leverage a transfer learning-based domain adaptation scheme, which shows promising results on the DF-W dataset. However, domain adaptation still requires a small number of deepfake videos from the target distribution/domain (DF-W in this case). Therefore, the attacker still has an upper hand, putting the defender in a difficult situation, unless we come up with defenses that can generalize better.
We also investigate the performance of the winning DNN model from Facebook's DFDC competition~\cite{dfdc_competition}. While this winning model outperforms the existing models (without domain adaptation), its performance on \wilddataset{} is still inadequate with an F1 score of 81\% and low precision of 71\%. 
\end{packed_itemize}

We release the \wilddataset{} dataset with the goal of enabling further work on deepfake detection. The \wilddataset{} dataset is available on our GitHub repository\footnote{\url{https://github.com/jmpu/webconf21-deepfakes-in-the-wild}}.

\section{Background}
\subsection{DeepFake Videos}
\begin{figure}[!t]
    \centering
    \includegraphics[width=1.0\columnwidth]{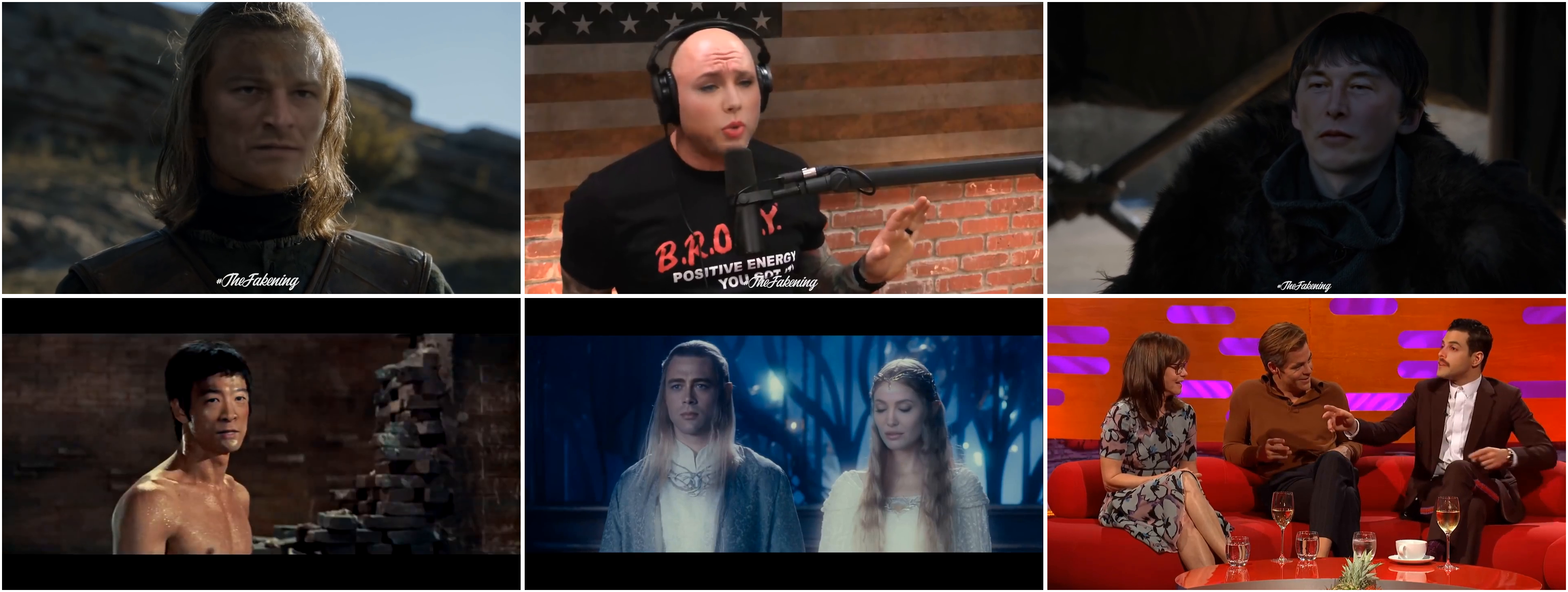}
    \caption{Frame samples of deepfake videos in the wild (YouTube).}
    \label{fig:deepfake-samples}
  \end{figure}

A deepfake video is popularly characterized as a video that has been manipulated using deep neural networks (DNNs), with the goal of simulating false visual appearances~\cite{deepfaketimit, chesney2019deep, li2018exposing, guera2018deepfake}. \textit{We focus on the most popular type of deepfakes found on the Internet, categorized as \textit{face-swapped} videos~\cite{agarwal2019protecting}.} This technique attempts to replace the face of an individual with that of another, while retaining the expression, pose, and background area of the image. Figure~\ref{fig:deepfake-samples} shows examples of such videos found in the wild that appear very convincing. \textit{In the rest of this work, we use the term ``deepfake video'' to refer to videos containing face-swapped content, and ``real video'' to refer to non-deepfake videos.} 

Deepfakes were first seen online in 2017, created by a Reddit.com user with the self-appointed name /u/DeepFakes~\cite{vice2017DeepFakearticle}. This user created deepfake videos of celebrities face-swapped into illicit videos, while only using publicly available data and the Tensorflow machine learning library. It was later revealed that the inspiration for their face-swap algorithm was an unsupervised image-translation work from NVIDIA~\cite{liu2017unsupervised}.
Since then, Internet communities have produced several deepfake generation tools (see Section~\ref{sec: generation_methods}) by leveraging state-of-the-art deep generative models proposed by the research community. Such models include Autoencoders~\cite{rumelhart1985learning}, Variational Autoencoders~\cite{kingma2014vae}, Convolutional Networks~\cite{krizhevsky2012imagenet}, and Generative Adversarial Networks (GANs)~\cite{goodfellow2014generative}. At the same time, the research community itself independently produced several variants of deepfake generation schemes (see Section~\ref{sec: generation_methods}). In this paper, we primarily focus on deepfakes produced by the Internet community, and appearing in the wild. 

\subsection{Methods for Generating Deepfakes}
~\label{sec: generation_methods}
We describe deepfake generation methods developed by both the Internet and research communities.
To aid our discussion, we adopt the following convention: the \textit{source} face is the face to be swapped in, and the \textit{target} face is the face that will be replaced.


\noindent\textbf{Methods by the Internet community.} The following methods 
developed by the Internet community, have been used to produce deepfakes we find on the Internet.

\noindent \textbf{(1) FaceSwap\footnote{https://faceswap.dev/}}: FaceSwap is an open-source tool created in 2017, 
using two deep autoencoders that
share the encoder module, but use different decoders. The two autoencoders are trained separately on the source and target faces to reconstruct each face from a latent representation.
The use of the shared encoder enforces a shared latent space, ensuring that the encoder disentangles facial identity from facial expression. To swap a face, the target face is fed into the shared encoder, and the source's decoder is used to decode the latent representation. The result is the swapped face, \ie with the source's face, but the target's facial expression and pose. The resulting face can then be spliced into the target image. 

\noindent \textbf{(2) DeepFaceLab (DFL)\footnote{\url{https://github.com/iperov/DeepFaceLab}}}: DFL is another open-source project created in 2018 as a fork of FaceSwap. On their website, it is claimed that ``More than 95\% of deepfake videos are created with DFL''. 
DFL boasts improvements over FaceSwap's model.
This includes the combined use of DSSIM~\cite{loza2006structural} and Mean Square Error reconstruction losses for autoencoder training. DFL offers 5 variants, each differing in terms of input and output face resolution (16x16 to 512x512), additional intermediate layers, architectural combinations of auto-encoder networks, and VRAM requirements. 

\noindent \textbf{(3) Zao\footnote{\url{https://www.zaoapp.net/}}}: Zao is an iOS mobile application created in 2018 by the Chinese software company, Momo.
Zao is closed source, and its methodology is unknown.

\noindent \textbf{(4) FakeApp\footnote{FakeApp is shut down, no link available}}: FakeApp, created in 2018, is owned and maintained by the online user DeepFakeapp. FakeApp's methodology is unknown, but is rumored to be based on the architecture of FaceSwap~\cite{fafsrumor}.

In our deepfake dataset introduced in Section~\ref{sec: wilddataset}, we find videos produced using the four methods described so far (\ie FaceSwap, DFL, Zao, and FakeApp). However, there are many videos in our dataset for which the generation method is unknown. For completeness, we briefly discuss some other methods developed by the Internet community.

\noindent \textbf{(5) Other methods:} A notable method is FaceSwap-GAN~\cite{faceswapgan}, created in 2018 as an open-source fork of FaceSwap. FaceSwap-GAN employs the same base DNN method as FaceSwap, with some improvements. The research community has used FaceSwap-GAN to produce the DeepFakeTIMIT~\cite{deepfaketimit} dataset.
Other methods include Dfaker~\cite{dfaker} and MyFakeApp~\cite{myfakeapp}, which also employ the same DNN architecture as FaceSwap. DeepFakesapp.online~\cite{deepfakeonline}, DeepFakes web $\beta$~\cite{deepfakeweb}, and DeepFake.me~\cite{deepfakeme} are websites that offer deepfake generation services behind paywalls (methodology is unknown) . Reflect~\cite{refecttech} and Doublicat~\cite{doublicat} are mobile applications with similarly unknown methodologies. 




\noindent\textbf{Methods by the research community.} Many deepfake generation methods have also been proposed by the research community. It is unclear if deepfakes found in the wild have directly used these methods. These methods are based on GANs, VAEs, CNNs, and encoder-decoder frameworks. This again shows that advances in generative modeling are enabling the creation of deepfakes. Methods include FSGAN~\cite{nirkin2019fsgan}, IPGAN~\cite{bao2018towards}, Fast Face Swap CNN~\cite{korshunova2017fast}, FaceShifter~\cite{li2019faceshifter}, FSNet~\cite{natsume2018fsnet}, RSGAN~\cite{natsume2018rsgan}, and DF-VAE~\cite{jiang2020deeperforensics}. 
\section{Deepfake Datasets}\label{sec: datasets}
In this section, we introduce our novel deepfake dataset called \wilddataset{}, containing deepfakes found in the wild (\ie produced by the Internet community). We also present details of existing deepfake datasets produced by the academic/industry research community.

\begin{table}[!t]
    \begin{tabular}{c|c|c|c|c}
    \hline
    \makecell{\bf Datasets} & \makecell{\bf \#Videos} & \makecell{\bf Total \\ \bf Frames} & \makecell{\bf Total \\ \bf Duration} & \makecell{\bf Avg. \\ \bf Duration}\\
    \hline
    \multicolumn{1}{l|}{\wilddataset{} YouTube} & 1,062  & 2.9M & 30h 1m 12s  & 1m 42s \\
    \multicolumn{1}{l|}{\wilddataset{} Bilibili} & 807 & 1.9M & 18h 48m 48s & 1m 24s   \\
    \hline
    \multicolumn{1}{l|}{\wilddataset{}} & 1,869 & 4.8M & 48h 50m 00s & 1m 34s
    \\
    \hline
    \end{tabular}
    \caption{Statistics for the DF-W Dataset.}
    \label{tab:in_the_wild_dataset_high_level_stats}
\end{table}

\vspace{-1ex}
\subsection{\wilddataset{}: Our New Deepfake Dataset}\label{sec: wilddataset}
We start by describing the data collection methodology used to build the \wilddataset{} dataset. We only focus on content that is ``safe for work'', \eg excluding pornographic, obscene or explicit material.\\
\noindent \textbf{Step 1: Searching and identifying potential deepfake videos.} To identify deepfake videos, we start our search from popular Internet platforms known to host or curate links to deepfake content. Our data sources are listed below. We use these different sources to build a list of potential deepfake videos. 

\noindent \textit{YouTube.} YouTube is known to host deepfake videos~\cite{youtube_has_deepfakes}. Our idea is to identify deepfake content creators or channels primarily uploading deepfake content. To identify such channels, we use the YouTube search feature, using keywords such as `deepfake' and `faceswap', amongst many others.\footnote{Other keywords/phrases included: synthetic videos, neurally generated videos, deep video, DeepFaceLab, FakeApp, fake video, artificially generated videos.} 
We add the YouTube channel URLs revealed by the first 10 pages of search results to our list. Beyond 10 pages we found content to be less relevant. To further expand the channel list, for each channel, we also add the related channels recommended by YouTube to our list. 
We also used Google Trends\footnote{\url{https://trends.google.com/trends/}} to identify more deepfake-related search queries, with the goal of finding more deepfake channels on YouTube. Starting from the initial set of 9 keywords, we discovered an additional 69 keywords. However, the new keywords did not yield any additional channels on YouTube.

\noindent \textit{Reddit.} Reddit.com hosts two popular deepfake discussion sub-forums, `GIF Fakes'\footnote{\url{https://www.reddit.com/r/GifFakes/}} and `SFWDeepFakes'.\footnote{\url{https://www.reddit.com/r/SFWdeepfakes/}} In these sub-forums, we found posts that linked to videos or channels, either created by the authors themselves, or by another content creator. We scraped all posts for URLs, and obtained 1,491 URLs. Out of the 1,491 URLs, 1,341 URLs pointed to YouTube videos and channels, and we follow the procedure described earlier in the YouTube section to add relevant YouTube channel URLs to our list.
The remaining 150 non-YouTube URLs led to platforms such as Reddit's own video-hosting service, Vimeo, and Imgur, all of which turned out to be duplicates of content we already identified on YouTube. We identified duplicates via manual examination of the content at these URLs. No non-video URLs were found amongst the 150 non-YouTube URLs.   


\noindent \textit{Bilibili~\cite{bibilibilink}}. Bilibili is a popular video sharing site in China, and is known to host videos from the Zao app~\cite{bilibili_has_deepfakes}. Zao is a free deepfake face-swapping app that can place your face into scenes from hundreds of movies and TV shows after uploading just a single photograph, and has gone viral in China~\cite{zaonews}. 
We again use the search feature using the keyword `zao'. Other keywords, \ie those used on the YouTube search, primarily revealed videos already available on YouTube, as well as non-relevant instructional and reaction videos.
We observed that nearly all uploading channels associated with videos in the search results, uploaded sparse amounts of deepfake content. Consequently, we iterated through every page of search results, and added the video URLs themselves to our list (in lieu of the channel URLs). We further manually clicked on these videos, to find related videos, but did not obtain any videos not already returned in the original search results.

Overall, in step 1, we shortlisted 194 YouTube channel URLs, likely hosting deepfake videos, and 1000 video URLs on Bilibili, likely containing deepfake content.

\noindent \textbf{Step 2: Filtering and downloading videos.} In this step, we first verify whether a video contains deepfake material, before downloading it. For verification purposes, we use a combination of manual and automated techniques to filter out non-deepfake videos. First, we used the YouTube API and the HTML source for Bilibili (which lacks an accessible API) after all JavaScript was loaded, to access metadata for each video. We then filtered out videos where the title or description contained no variations of the search keywords originally used in Step 1. Each remaining video (from both YouTube and Bilibili) was then manually verified as to whether it contained face-swapped content. This was done by looking for statements in the title, description or comments claiming the video to be a deepfake, and by also looking for facial flickering, inhuman feature distortion, warping and obvious lighting mismatches.\footnote{We did find cases where it was hard to make a decision just based on content (\eg for high-quality deepfakes with no visible artifacts). In such cases, we relied on statements in the title, and description.} 


Lastly, we downloaded all the videos (verified to be deepfakes) at the highest available resolution, using the open-source tool youtube-dl\footnote{\url{https://github.com/ytdl-org/youtube-dl}}. 
Videos that were not made available on YouTube to the author's home region were not downloaded.
All our measurement efforts were conducted between 2020-02-01 and 2020-03-01.

\begin{table*}[!ht]
    \begin{tabular}{c|c|c|c|c|c|c}
    \hline
    \textbf{Dataset}  & \textbf{\makecell{Release \\Date}} & \makecell{\bf \#Real Videos} & \bf Real Source & \textbf{\#Fake Videos}  & \textbf{\makecell{Generation \\Method}} & \textbf{\makecell{Avg. \\Duration}}  \\
    \hline
    \multicolumn{1}{l|}{UADFV~\cite{UADFV}} & 2018.11 & 49& YouTube    & 49 & FakeApp & 11.55s\\
    \hline
    \multicolumn{1}{l|}{DeepFakeTIMIT~\cite{deepfaketimit} } & 2018.12  & 320 & VidTIMIT dataset~\cite{vidtimit} & 320 & FaceSwap-GAN & 4.40s\\
    \hline
    \multicolumn{1}{l|}{FaceForensics++~\cite{rossler2019faceforensics++}} & 2019.01 & 1,000 & YouTube & 1,000 & FaceSwap & 18.72s\\
    \hline
    \multicolumn{1}{l|}{DFD~\cite{googledfd}} & 2019.09 & 363 & Self-recording & 3,068 & Unknown method & 30.30s\\
    \hline
    \multicolumn{1}{l|}{Celeb-DF~\cite{li2019celeb}} & 2019.11 & 590 &YouTube & 5,639 & Unknown method & 12.51s\\
    \hline
    \multicolumn{1}{l|}{DFDC~\cite{dolhansky2020deepfake}}& 2019.12 & 23,654 & Self-recording  & 104,500 & 8 generation methods~\footnotemark & 10.02s \\
    \hline
    \end{tabular}
    \vspace{1ex}
    \caption{Statistics of deepfake datasets produced by the research community.}
    \label{tab:academic_dataset_high_level_stats}
    \vspace{-3ex}
\end{table*}
\noindent \textbf{\wilddataset{} dataset.} Using the above methodology, we prepared the \wilddataset{} dataset containing a total of 1,869 deepfake videos, of which 1,062 are from YouTube, and 807 are from Bilibili. Dataset statistics are shown in Table~\ref{tab:in_the_wild_dataset_high_level_stats}. In total, our dataset contains over 48 hours of video content, with over 4.8 million frames (extracted at the native frame rate). 
Video resolutions range from (360 x 360) to a high resolution of (2560 x 1080).
Additionally, we collected the following metadata for each video using the YouTube API and Bilibili HTML source: publishing date, number of views, number of subscribers for the channel (content creator), and the total number of videos associated with the video's channel.

We note that \textit{concurrent work} by Zi et al.~\cite{zi2020wilddeepfake} also introduces a real-world deepfake dataset, comprising face sequences extracted from 707 deepfake videos collected from the internet. To the best of our knowledge, \wilddataset{} is the \textit{largest} collection of deepfake videos in the wild.

\vspace{-1ex}
\subsection{Research Community Datasets}\label{sec: research_datasets}
Researchers have released a variety of datasets to study deepfake video detection. We use these datasets in our study, and compare them with our \wilddataset{} dataset in Section~\ref{sec: wilddataset}.
Dataset statistics are in Table~\ref{tab:academic_dataset_high_level_stats}.  
Each dataset varies across different dimensions, \eg the number of videos, generation methods used, how real videos are obtained, video content, and quality. Video resolutions range widely from 234x268 to 1920x1080.
Notably, in datasets released by researchers, \eg UADFV, FaceForensics++ and Celeb-DF, real videos are collected from YouTube or an off-the-shelf video set, based on which they generate deepfake videos. While in datasets released by industry, \eg DFD and DFDC,  real videos are recorded by paid actors. Most recently, Jiang et al. proposed the DeeperForensics dataset in 2020~\cite{jiang2020deeperforensics}. However, due to the timing of the release, we are unable to include it in our analysis. According to prior work~\cite{li2019celeb}, UADFV, DeepFakeTIMIT, and FaceForensics++ are considered to be the first generation datasets. These datasets have lower-quality synthesized faces, with some visible artifacts from the generation process, and lower pose variations. The second generation datasets include DFD, Celeb-DF, and DFDC, and they improve on many of the limitations of the first generation datasets.

\noindent \textbf{\researchdataset{} dataset.} We additionally prepare a dataset representing all the research community datasets, by sampling deepfake videos from the 6 datasets, and call it the \researchdataset{} dataset. This simplifies analysis, enabling us to compare detection performance on the \wilddataset{} dataset, with a single deepfake dataset representing the research community. 
We strive to keep the number of deepfake videos similar to that in the \wilddataset{} dataset, and therefore sample 500 random videos (or less if fewer are available) from the deepfake and real classes of each research community dataset. The \researchdataset{} dataset contains a total of 2,369 deepfake videos, and 2,316 real videos.

\vspace{-1ex}
\section{Analyzing In-The-Wild Deepfakes}
\label{sec:analyze-deepfakes}
Here our goal is to provide a deeper understanding of the \wilddataset{} dataset, and how it compares with existing datasets from the research community (\researchdataset{}). We also discuss the characteristics of the \wilddataset{} dataset that raises implications for a defender.




\begin{figure*}[!ht]
\centering
\begin{subfigure}[t]{0.24\textwidth}
\includegraphics[width=1.0\linewidth]{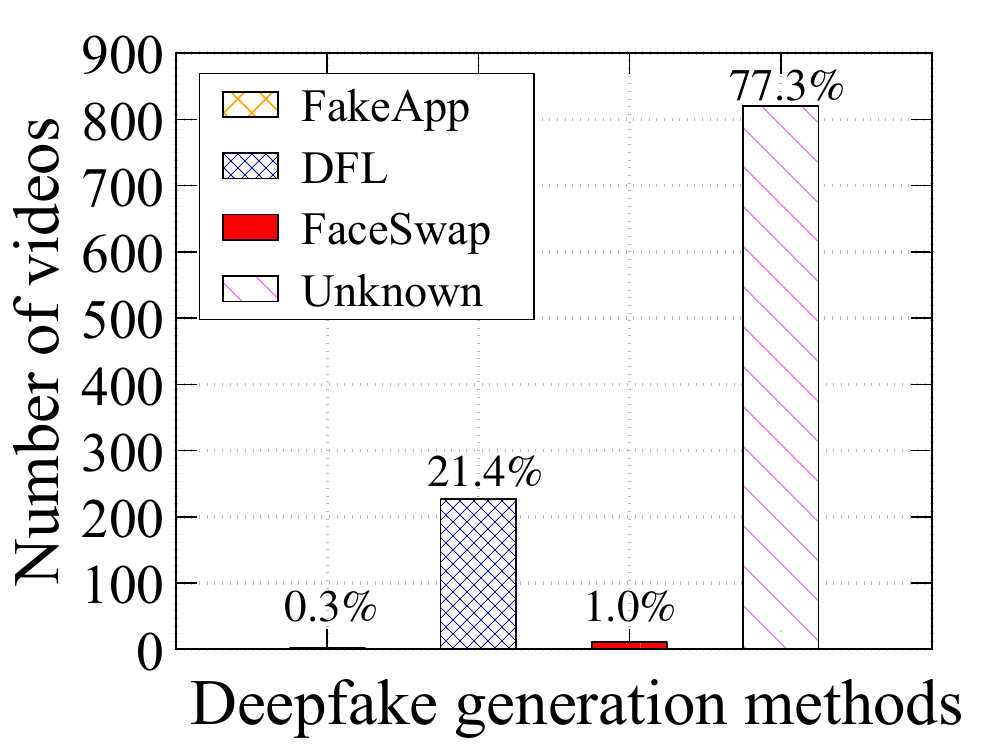}\caption{}
\label{fig:methods_pie}
\end{subfigure}
\hfill
\begin{subfigure}[t]{0.24\textwidth}
\includegraphics[width=1.0\linewidth]{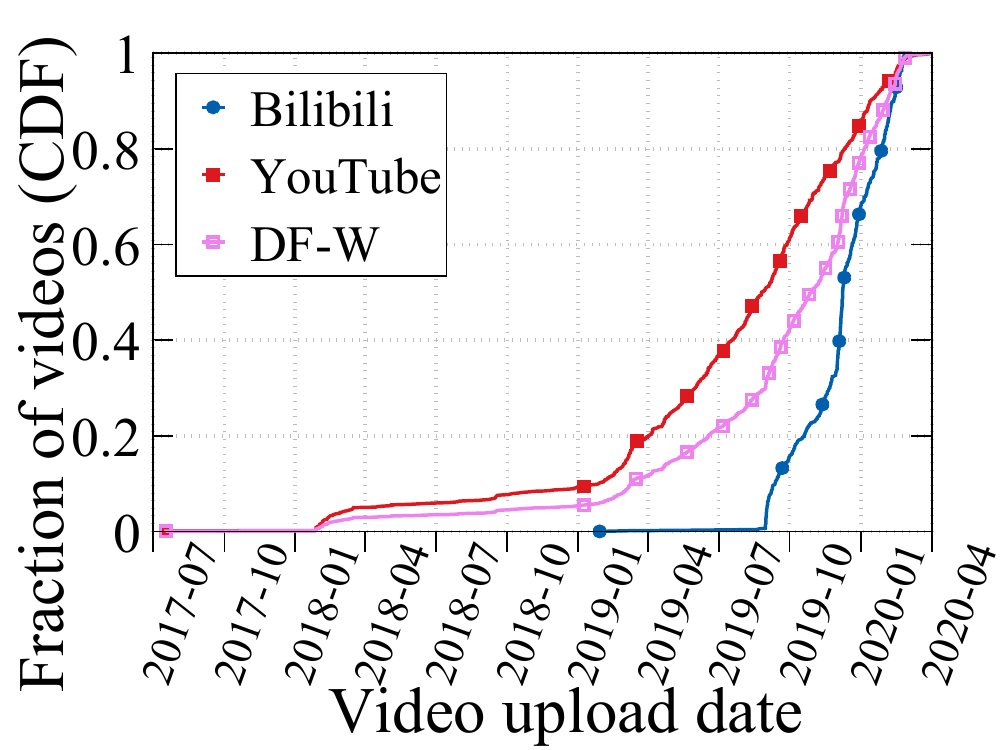}
\caption{}
\label{fig:video-growth-over-time}
\end{subfigure}
\hfill
\begin{subfigure}[t]{0.24\textwidth}
\includegraphics[width=1.0\linewidth]{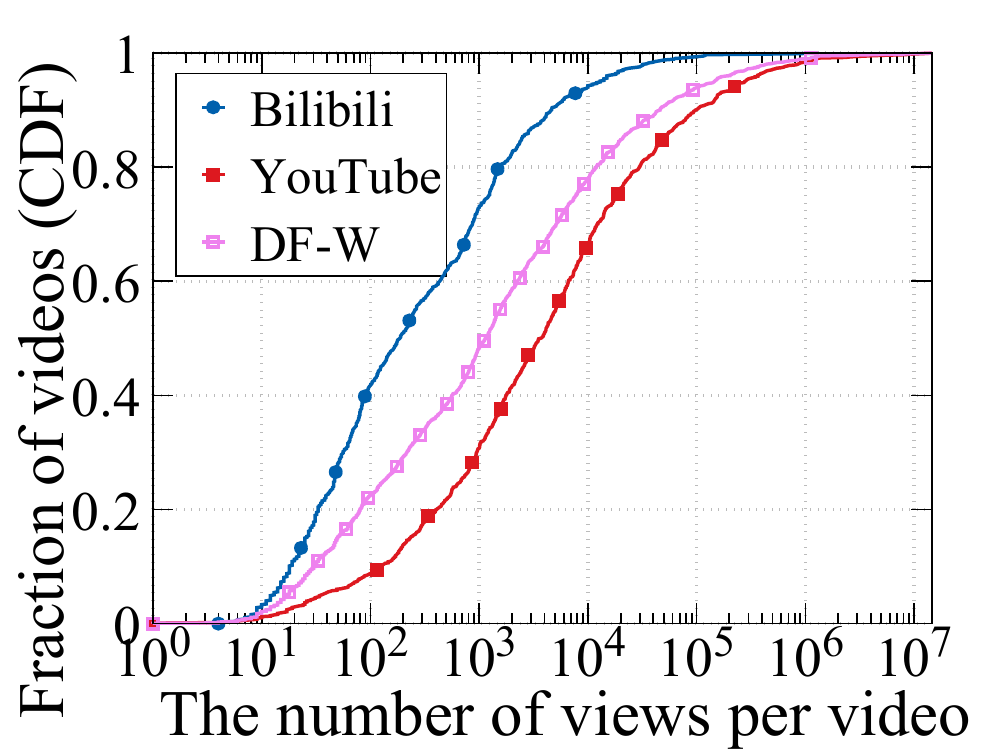}
 \centering
 \caption{}
\label{fig::video-views}
\end{subfigure}
\hfill
\begin{subfigure}[t]{0.24\textwidth}
  \includegraphics[width=1.0\linewidth]{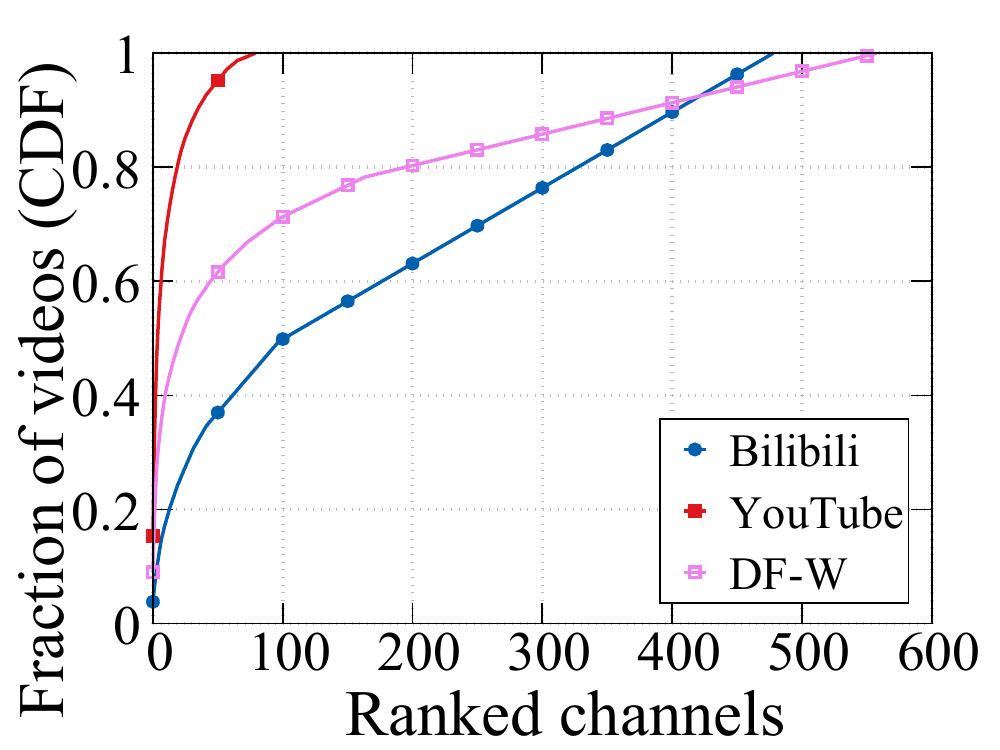}
\caption{}
\label{fig:rank-plot-channels}
\end{subfigure}
\vspace{-1ex}
\caption{(a) Distribution of generation tools used for creating \wilddataset{} YouTube (b) CDF of \wilddataset{} video upload dates (c) distribution of generation tools used for creating \wilddataset{} YouTube (d) CDF of number of \wilddataset{} videos contributed by channels.}
\vspace{-3ex}
\end{figure*}

\vspace{-2ex}
\subsection{DF-W vs Research Community Datasets}\label{sec: metadata_dfw_vs_dfr}

\noindent \textbf{Deepfake generation methods.} We examine the deepfake generation methods used by the different datasets. To determine the generation methods for \wilddataset{} videos, we manually examine the title and description of each video for any mention of a particular tool or method. The description often contains a link to the tool. 
For instance, the description of the `David Bowie - I Believe [DeepFake]' video states: `Software used: \url{https://github.com/iperov/DeepFaceLab/}'. If the video meta-data did not state any method, we mark the method as \textit{unknown}.
Using this method, we find that all the videos from Bilibili are produced using the Zao iOS application. For YouTube, we were able to obtain the generation method label for
241 out of 1,062 videos (22.71\%), while the remaining are marked as \textit{unknown}. Figure~\ref{fig:methods_pie} shows the distribution of the three deepfake methods that we found on YouTube: FakeApp, FaceSwap, and DFL. 
DFL makes up the vast majority (94.2\%) of videos for which we have a known method. 


Among the research datasets, generation methods are unspecified in Celeb-DF and DFD. FaceSwap, FakeApp, and FaceSwap-GAN were used to generate FaceForensics++, UADFV, and DeepFakeTIMIT, respectively. DFDC was generated using 8 custom autoencoder-based, GAN-based, and non-neural methods. Interestingly, DFL is not used by any of the existing research community datasets, but is a popular generation method in the wild. \textit{This indicates a mismatch in the methods used by the research community and the Internet community.} Using methods from the wild would help to build a more representative deepfake dataset.

\noindent \textit{Can we automatically infer the generation method for a video?} For a large fraction (77\%) of DF-W YouTube videos, the generation method is unknown. A defender with knowledge of the generation methods can create effective targeted defenses. Therefore, we propose to infer the generation scheme for the videos (with unknown methods) using a DNN-based method. We focus on the generation method, DFL, mainly for two reasons---it is the only method where we have sufficient data to conduct our analysis, and DFL, themselves claim to be the most popular method (Section~\ref{sec: generation_methods}). Our scheme to fingerprint the generation method aims to determine whether a video is created using DFL or a non-DFL method (\ie some other method). We first train this classifier using a labeled dataset of DFL and non-DFL videos, and evaluate its performance, before applying it to the videos with unknown generation schemes.

To build the fingerprinting scheme, we leverage the DNN model from Yu et al.~\cite{yu2019attributing}, which was proposed as a method to fingerprint the GAN model, given a GAN generated image. In our setting, the input to the classifier is a face extracted from a frame, which is then classified as belonging to the DFL or the non-DFL class.
We randomly sample 200 DFL videos (from DF-W), and 200 non-DFL videos. Non-DFL videos are sampled from DF-W and research community datasets, and covers 4 generation methods, \ie FakeApp, FaceSwap-GAN, FaceSwap, and Zao. Next, to identify and extract fake faces from each video, we use the deepfake detection scheme called CapsuleForensics~\cite{nguyen2019use}.
\footnotetext{There are 8 deepfake generation methods used in the DFDC dataset. Examples include DF-128, DF-256, MM/NN, NTH, FSGAN, and StyleGAN~\cite{dolhansky2020deepfake}.}
\footnote{In DF-W videos, not all faces/frames contain fake content.} CapsuleForensics provides a probability score for a face being fake, and we only choose faces with a probability score higher than $0.7$. We randomly sample 20k, 4k, and 4k faces (with balanced ratio of fake and real faces) using this criteria for training, validation, and testing, respectively. We use hyperparameters recommended by Yu et al., and also apply dropout (prob=0.5) to the last two dense layers. On tuning the decision threshold for high precision, our final model achieves a 90\% precision and 70\% recall on the testing set for the DFL class. We then run this trained model on 821 source-unknown videos from \wilddataset{} YouTube. A video is flagged as DFL-generated if more than 50\% faces of the video are flagged as DFL-generated. Finally, 243 videos (29.6\%) are classified as DFL-generated out of the 821 unknown videos in \wilddataset{} YouTube. \textit{Our analysis suggests that DFL is indeed a popular tool in the wild, and should be considered by the academic community to create deepfake datasets.}

\noindent \textbf{Content analysis.} We describe key differences between the deepfake content in the \wilddataset{} dataset, and the research community datasets. \textit{First,} for all research community datasets, every frame contains some fake content (\ie a swapped face). In fact, many existing deepfake detection schemes (\eg CapsuleForensics~\cite{nguyen2019use}, Xception~\cite{rossler2019faceforensics++}, FWA~\cite{li2018exposing}, MesoNet~\cite{afchar2018mesonet}, VA, ~\cite{matern2019exploiting}, and Multi-Task~\cite{nguyen2019multi}) are evaluated under the assumption that every frame --- with a detectable face --- in a deepfake video contains fake content. However, this is not the case for videos in the \wilddataset{} dataset. We find several cases where only a fraction of the frames contains fake content, including cases where majority of frames are clean.\footnote{It is hard to quantify using exact numbers --- requires manual examination of millions of frames.} This has implications for the design of detection schemes (Section~\ref{sec: experimental_setup}), because a large number of clean frames (\ie without fake content) can increase the risk of false negatives. 

\textit{Second,} we examine the number of faces in a frame, and the number of fake faces in a frame. Some existing detection schemes (\eg CapsuleForensics, Xception, Multi-Task~\cite{nguyen2019multi}) assume that each frame only contains a single face which is faked. Such methods would need to be adapted and calibrated to handle cases where there are multiple faces in a frame, of which all or only a portion of faces are fake. From manual inspection, we observe three variations of deepfakes in the \wilddataset{} dataset: (1) single face that is fake, (2)
multiple faces with a single fake face, and 3) multiple faces and multiple
fake faces. 
In contrast, some research community datasets, UADFV and DeepFakeTIMIT fall entirely in the first category. To further quantify this, we use a CNN-based face detection model called dlib~\cite{king2009dlib} to identify the number of faces in a video. We approximately estimate the number of faces in a video, as the maximum number of faces detected in any frame in the video. We find that nearly 26\% of \wilddataset{} videos contain more than a single face, whereas, for all 6 research community datasets (in Section~\ref{sec: research_datasets}) 92\% to 100\% of videos contain only a single face.

\textit{Third,} we observe that \wilddataset{} videos are longer in duration than research datasets. In particular, the longest duration
among research datasets is less than $100$ seconds; however, nearly 32\% of videos in \wilddataset{} have a duration longer than $100$ seconds.
In UADFV and DFDC, all the videos are shorter than 10s, and in CelebDF all are shorter than 20s. In comparison, the videos in \wilddataset{} range from 10s to 500s. Longer duration, combined with variations of deepfake content (described earlier), can lead to more false negatives while detecting deepfakes, \eg video has large number of clean frames. 

\vspace{-2ex}
\subsection{Growth, Popularity, and Creators of DF-W Videos}
\noindent \textbf{Content growth.} Figure~\ref{fig:video-growth-over-time} shows the distribution of the upload dates of \wilddataset{} videos. 
Bilibili has a sharp increase in uploads in September 2019, when the Zao app went viral in China. For YouTube, the number of uploaded deepfake videos saw a slight bump in January 2018 with the original release of FakeApp, but the largest jump happened at the beginning of 2019 when DFL and FaceSwap became popular open source alternatives to FakeApp. \textit{The steady recent growth of deepfake content suggests the urgent need to build effective defenses against misuse of deepfakes.}

\noindent \textbf{Popularity.} In both YouTube, and Bilibili, there are videos that are hugely popular. There are 31, and 2 videos in YouTube, and Bilibili, respectively, with over 500,000 views. There are more popular videos on YouTube than Bilibili. Nearly 69\%, and 27\% of videos in YouTube, and Bilibili, respectively, received over 1K views. Distribution of the number of views per video is shown in Figure~\ref{fig::video-views}. Moreover, when we investigate the ratio of the number of views to the number of subscribers of the corresponding channel, we find that 60\% and 89\% of videos on YouTube and Bilibili, respectively, have a ratio higher than 1. This implies that deepfake content easily and often spreads beyond the audience of its channel. Our analysis shows that deepfake content can reach a large audience, thus benefiting an attacker creating misleading deepfakes.

\noindent \textbf{Content creators.} Here we study the channels or content creators that upload the deepfake videos. There are 80 channels for YouTube, and 477 channels for Bilibili. Figure~\ref{fig:rank-plot-channels} shows the ranked plot of channels, and the cumulative fraction of videos associated with them. For YouTube, the top 16 channels (covering 20\% of channels) account for  75\% of the videos. This suggests that there are a small number of creators/channels on YouTube who have the resources to create a large number of deepfake videos. However, we observe a long tail  distribution for Bilibili, where the top 95 channels (covering 20\% of channels) only account for 49\% of the videos. In Bilibili, over 90\% of channels only upload one or two deepfake videos. This suggests that defense efforts need to be prepared to tackle both resourceful deepfake content creators who can churn out large number of deepfake videos, as well as a large number of creators/users who may each produce only a few deepfake videos. The generation methods used by these creators could be different, thus potentially making defense efforts harder.

\vspace{-1ex}
\section{Detecting DeepFakes}\label{sec:detection}
We evaluate the performance of existing detection methods on \wilddataset{} and \researchdataset{} datasets, and also evaluate approaches for improving classification performance.

\vspace{-2ex}
\subsection{Experimental Setup}\label{sec: experimental_setup}
\noindent \textbf{Detection methodology and performance metrics.}
Most prior work has framed the problem as a frame-level
binary classification task, where the probability score of a frame being fake is computed using the extracted face as input. 
Using this
formulation, classifiers have portrayed impressive accuracy, and ROC AUC performance scores at the frame
level~\cite{li2018exposing, nguyen2019use, rossler2019faceforensics++, li2019celeb, matern2019exploiting}. 
Other prior work used a video-level decision metric by assigning the
video probability score (of being fake) as the \textit{average} of all frame probability scores~\cite{afchar2018mesonet, nguyen2019use}.

The above metrics work well for existing research community datasets for the following reasons: (1) Existing datasets (see Section~\ref{sec: research_datasets}) contain deepfake videos that are entirely fake, \ie every frame with a face has been manipulated. (2) In existing datasets, most (92\%--100\%) deepfake videos contain at most one face per frame (see Section~\ref{sec: metadata_dfw_vs_dfr}). However, these assumptions do not hold for the DF-W videos. In Section~\ref{sec: metadata_dfw_vs_dfr}, we discuss that \wilddataset{} videos can contain multiple faces per frame, not all of which are  fake. Also, not all frames necessarily contain a fake face.

Therefore, we need to modify the decision metric to incorporate scores for \textit{multiple} faces. Further, since we lack frame level ground-truth, our
evaluation metric needs to be at the video level. However, averaging frame scores to obtain
the video score, as used by prior work is problematic since a large portion of
the video can contain real, unmodified frames. 

To tackle the above challenges, we propose a modified version of a video-level
decision heuristic, originally introduced by Li et al~\cite{li2018exposing}. 
Li et al. propose to compute the probability score for a
given video being fake by averaging the top $F_p$ percentile of frame probability scores: $P(v) = \dfrac{1}{n}\sum_{i=F_{p}\%}^{100\%} P(x_{i})$, where the percentile value $F_p$ is a decision parameter. While Li et al. assume one face per frame, we compute the frame-level probability score in the above equation, as follows:
$P(x_i) = max(P_{f_1}, P_{f_2}, P_{f_3}, \dots)$ where $x_i$ refers to the $i^{th}$ frame, and $P_{f_j}$ refers to the probability score for the $j^{th}$ face in the frame.  

Li et al. used an $F_{p}$ value of 33\%. 
To provide better insight into the current performance limits for deepfake detection, we adopt the F1-score metric, and sweep the parameter space for the optimal $F_p$. More specifically, for each given classifier and testing set configuration, we compute detection performance using the value of $F_p$ that allows for the best-attainable F1 score. Computing the best-attainable F1 score, helps us understand the upper bound on performance.

\noindent \textbf{Supervised methods.} Supervised deepfake detection methods refer to those that use a labeled dataset of deepfake and real videos to train a classifier. To evaluate the current state-of-the-art, we evaluate those supervised methods that have been proposed in peer-reviewed work, have released code, and pre-trained models.

\noindent \textbf{(1) CapsuleForensics~\cite{nguyen2019use}:} Nguyen et al. proposed CapsuleForensics which
employs a VGG network~\cite{simonyan2014very}, and a Capsule network~\cite{hinton2011transforming}. CapsuleForensics is trained and tested on the FaceForensics++ dataset.
More specifically, the training data comprises faces extracted from frames of 2,160 deepfake videos generated with FaceSwap, from 5,320 non-neurally manipulated videos, and from 2,160 real YouTube videos.
Their paper reports detection performance at the frame-level with a 93\% accuracy, and 92\% accuracy at the video-level (using overall frame-probability averaging). 
    
\noindent \textbf{(2) Xception~\cite{rossler2019faceforensics++}:} Rossler et al. proposed the Xception method, a CNN-based model that is pre-trained on Imagenet, and trained and tested on an updated version of the FaceForensics++ dataset. 
In addition to all the videos in the version used by CapsuleForensics, this updated dataset also comprises faces extracted from frames of 2,160 manipulated videos generated using the neural expression-manipulation tool, NeuralTextures~\cite{thies2019deferred}.
They report a 99\% frame level accuracy.

\noindent \textbf{(3) MesoNet~\cite{afchar2018mesonet}:} Afchar et al. proposed MesoNet, a CNN-based model with two configurations, Meso4 and MesoInception4. We choose Meso4, as both models exhibit similar performance. Meso4 is trained and tested on a dataset containing a small number of deepfake videos collected from YouTube, and real images collected from the Internet. More specifically, the training data comprises faces extracted from frames of 175 deepfake videos, and 7,250 real faces collected from ``various Internet sources'', chosen such that their resolution distribution matches that of faces from the deepfake videos.
They report a frame level accuracy of 89\%, and a video level accuracy (using overall frame-probability averaging) of 97\%.


\noindent \textbf{(4) Multi-Task~\cite{nguyen2019multi}:} Nguyen et al. proposed Multi-Task, based on a multi-output autoencoder. Multi-Task is trained and tested on a subset of the FaceForensics++ dataset. 
The training data includes faces extracted from frames of 704 non-neurally manipulated videos, and 704 real YouTube videos. 
Multi-Task is designed for generic facial-manipulation detection, for which they report frame-level testing accuracies of upto 92\%. 
    
\noindent \textbf{(5) VA~\cite{matern2019exploiting}:} Matern et al. proposed VA,
a method using either a multi-layer perceptron (VA-MLP), or a logistic regression classifier (VA-LReg). Both models show similar performance, and we choose the VA-MLP classifier. VA-MLP is trained and tested on a dataset containing YouTube videos with explicitly demarcated `side-by-side' deepfake and real content. In such videos, the content creator puts the original and the deepfake versions of the video horizontally next to each other, allowing the viewer to watch the entire video with both views. The dataset also contains normal deepfake videos collected from YouTube. More specifically, the training data comprises faces with eyes and mouth open, extracted from the `fake side' of frames of 4 `side-by-side' deepfake videos and from 3 normal deepfake videos. The real faces are similarly extracted from the real side of the frames of the same 4 `side-by-side' videos.
They report 85\% frame-level ROC AUC score.



\begin{table}[!t]
    \centering
    \small
    \begin{tabular}{c|c|c|c|c|c|c|c|c|c|c|c|c|c|c|c|c|c|c|c|c|c}
        \hline
        \multicolumn{22}{c}{\bf \wilddataset{}} \\
        \hline
        \multicolumn{1}{c|}{\bf Class} & \multicolumn{6}{c|}{\bf \wilddataset{}} & \multicolumn{6}{c|}{\bf \wilddataset{} YouTube} & \multicolumn{9}{c}{\bf \wilddataset{} Bilibili}\\
        \hline
        Fake & \multicolumn{6}{c|}{1,864} & \multicolumn{6}{c|}{1,057~\footnotemark} & \multicolumn{9}{c}{807} \\
        Real & \multicolumn{6}{c|}{1,864} & \multicolumn{6}{c|}{1,057} & \multicolumn{9}{c}{806} \\
        \hline
        \multicolumn{22}{c}{\bf \researchdataset{}} \\
        \hline
        \bf Class & \multicolumn{3}{c|}{\thead{\bf \researchdataset{}}} &
        \multicolumn{3}{c|}{\thead{\bf \researchdataset{} \\ \bf FFPP}} &
        \multicolumn{3}{c|}{\thead{\bf \researchdataset{} \\ \bf Celeb-DF}} &
        \multicolumn{3}{c|}{\thead{\bf \researchdataset{} \\ \bf DFDC}} &
        \multicolumn{3}{c|}{\thead{\bf \researchdataset{} \\ \bf DFD}} &
        \multicolumn{3}{c|}{\thead{\bf \researchdataset{} \\ \bf Dtimit}} &
        \multicolumn{3}{c}{\thead{\bf \researchdataset{} \\ \bf UADFV}} \\
        \cline{2-22}
        \hline
        Fake & \multicolumn{3}{c|}{2,369} & \multicolumn{3}{c|}{500} & \multicolumn{3}{c|}{500} & \multicolumn{3}{c|}{500} & \multicolumn{3}{c|}{500} & \multicolumn{3}{c|}{320} & \multicolumn{3}{c}{49} \\
        Real & \multicolumn{3}{c|}{2,316} & \multicolumn{3}{c|}{501} & \multicolumn{3}{c|}{501} & \multicolumn{3}{c|}{501} & \multicolumn{3}{c|}{501} & \multicolumn{3}{c|}{320} & \multicolumn{3}{c}{50} \\
        \hline
    \end{tabular}
    \vspace{1ex}
    \caption{Testing set sizes for evaluating detection performance on \wilddataset{} and \researchdataset{} datasets.}
    \label{tab:detection_testing_set_sizes}
\end{table} 

\footnotetext{For 5 videos of \wilddataset{} YouTube (1,062 videos), no faces were detected by the face detector.}

\noindent \textbf{Unsupervised methods.}
Unsupervised deepfake detection methods only use real videos/images to train a classifier.
We focus on the following two unsupervised methods.

\noindent \textbf{(1) FWA~\cite{li2018exposing}:} Li et al. proposed the FWA method, a CNN-based model.
FWA is trained on a dataset containing real faces, intentionally warped to simulate deepfake artifacts, and unwarped real faces collected from the Internet. Warping is achieved by rescaling a real face to a random scale, smoothing with a $5x5$ Gaussian blur, and then affine warping back to the original scale. The training data comprises 24,442 warped real faces, and 24,442 unwarped real faces. They report upto 99.9\% frame-level ROC AUC.

\noindent \textbf{(2) DSP-FWA\footnote{\url{https://github.com/danmohaha/DSP-FWA}}:} Li et al. proposed the DSP-FWA method in 2019. This method employs a CNN that improves upon FWA. DSP-FWA is trained on data identical to that of FWA. No reported performance numbers are available.

\noindent\textit{For all methods (supervised and unsupervised), we directly use the pre-trained models provided by the model developers.}

\noindent \textbf{Our testing datasets.} 
 We present detection performance on the deepfake videos in the \wilddataset{} dataset as a whole, as well as separately on the individual YouTube and Bilibili partitions. Detection performance on the \researchdataset{} dataset is also presented as a whole, as well as separately on each of the 6 individual research datasets (Section~\ref{sec: research_datasets}). For all datasets, we downsampled the larger class to maintain a balanced ratio of real and deepfake videos. 
The final testing set sizes for the deepfake and real classes are listed in Table~\ref{tab:detection_testing_set_sizes}.

\vspace{-1ex}
\subsection{Performance of Existing Detection Schemes} \label{sec:detection-performance}
In this section, we evaluate performance of existing detection methods. Our objective is to understand the best-attainable F1 score (in detecting fake videos) for each configuration. This implies that our results are optimistic, \ie in favor of the tested methods. In practice, the methods will perform sub-optimally if configured with a sub-optimal $F_p$ threshold. Table~\ref{tab:detection_performances} presents our results.

\begin{table*}[h]
    \centering
    \begin{tabular}{l|>{\columncolor[rgb]{0.83,0.83,0.83}}c|c|c|>{\columncolor[rgb]{0.83,0.83,0.83}}c|c|c|>{\columncolor[rgb]{0.83,0.83,0.83}}c|c|c|>{\columncolor[rgb]{0.83,0.83,0.83}}c|c|c|>{\columncolor[rgb]{0.83,0.83,0.83}}c|c|c|>{\columncolor[rgb]{0.83,0.83,0.83}}c|c|c|>{\columncolor[rgb]{0.83,0.83,0.83}}c|c|c}
    \hline
    \multirow{3}{*}{\bf Dataset} & \multicolumn{15}{c|}{\bf Supervised} & \multicolumn{6}{c}{\bf Unsupervised} \\
    \cline{2-22}
    & \multicolumn{3}{c|}{\bf Capsule} &
    \multicolumn{3}{c|}{\bf Xception} & \multicolumn{3}{c|}{\bf Mesonet} &  \multicolumn{3}{c|}{\bf Multi-Task} &  \multicolumn{3}{c|}{\bf VA} &  \multicolumn{3}{c|}{\bf DSP-FWA} &  \multicolumn{3}{c}{\bf FWA}  \\
    \cline{2-22}
    & \bf F1 & \bf P & \bf R & \bf F1 & \bf P & \bf R & \bf F1 & \bf P & \bf R & \bf F1 & \bf P & \bf R & \bf F1 & \bf P & \bf R & \bf F1 & \bf P & \bf R & \bf F1 & \bf P & \bf R \\
    \hline
    \wilddataset{} & \underline{\bf 77} & \bf 69 & \bf 86 & \bf 66 & \bf 67 & \bf 66 & \bf 74 & \bf 64 & \bf 87 & \bf 67 & \bf 51 & \bf 99 & \bf 67 & \bf 52 & \bf 95 & \bf 76 & \bf 63 & \bf 95 & \bf 73 & \bf 60 & \bf 93\\
    \wilddataset{} YouTube & 74 & 69 & 80 & 68 & 67 & 68 & 73 & 65 & 84 & 67 & 51 & 99 & 68 & 53 & 95 & \underline{79} & 69 & 93 & 73 & 61 & 90\\
    \wilddataset{} Bilibili & \underline{78} & 69 & 90 & 65 & 66 & 63 & 76 & 66 & 89 & 68 & 51 & 100 & 67 & 53 & 91 & 75 & 61 & 96 & 74 & 60 & 97\\
    \hline
    \researchdataset{} & \bf 66 & \bf 57 & \bf 78 & \bf 60 & \bf 80 & \bf 48 & \bf 55 & \bf 51 & \bf 58 & \bf 67 & \bf 51 & \bf 100 & \bf 61 & \bf 49 & \bf 81 & \bf \underline{72} & \bf 61 & \bf 86 & \bf 67 & \bf 53 & \bf 93\\
    \researchdataset{} FFPP & 95 & 95 & 96 & \underline{100} & 100 & 100 & 67 & 58 & 79 & 66 & 50 & 99 & 58 & 46 & 77 & 93 & 88 & 98 & 81 & 75 & 88\\
    \researchdataset{} Celeb-DF & 57 & 52 & 65 & 38 & 48 & 31 & 50 & 49 & 52 & 67 & 50 & 100 & 62 & 49 & 84 & 61 & 51 & 75 & \underline{70} & 55 & 96\\
    \researchdataset{} DFDC & 63 & 58 & 68 &42 & 51 & 35 & 54 & 51 & 58 & 68 & 51 & 100 & 65 & 51 & 89 & 82 & 79 & 85 & \underline{83} & 76 & 91\\
    \researchdataset{} DFD & 77 & 82 & 72 & \underline{84} & 94 & 76 & 67 & 59 & 79 & 67 & 50 & 100 & 58 & 47 & 76 & 76 & 72 & 81 & 57 & 46 & 76\\
    \researchdataset{} DTimit & 53 & 51 & 56 & 0 & 0 & 0 & 11 & 13 & 9 & 67 & 51 & 99 & 62 & 53 & 73 & \underline{78} & 67 & 93 & 70 & 55 & 96\\
    \researchdataset{} UADFV & 67 & 80 & 57 & \underline{94} & 98 & 90 & 37 & 40 & 35 & 67 & 50 & 100 & 61 & 51 & 76 & 86 & 91 & 82 & 83 & 74 & 94\\
    \hline
    \end{tabular}
    \caption{Best attainable detection performances (F1) of various classifiers on the \wilddataset{} and \researchdataset{} datasets, with corresponding precision (P) and recall (R) scores. Best attainable detection performance (F1) for each dataset is underlined. }
    \label{tab:detection_performances}
\end{table*}

\noindent \textbf{Detection performance on the \wilddataset{} dataset.}\label{sec:detection_performances} The first row in Table~\ref{tab:detection_performances} presents (best-attainable) detection performance for each classifier on \wilddataset{}. The performance of all classifiers is poor --- all F1 scores are below 77\%, going as low as 66\%. Moreover, all precision scores are below 69\%, suggesting many false positives. \textit{This indicates that these classifiers fail to generalize to a variety of real-world deepfake videos.} 

The best-performing supervised classifier is CapsuleForensics, with an F1 score of 77\%, followed by MesoNet, with an F1 score of 74\%. The remaining supervised classifiers exhibit worse performance, with F1 scores below 70\%. Multi-Task performs particularly poorly. We observe it produces the same classification decision of `fake' for nearly all inputs, confirmed by effectively random precision and perfect recall.

The best-performing unsupervised classifier on \wilddataset{} is DSP-FWA, with an F1 score of 76\%, followed by FWA, with a F1 score of 73\%. Performance of unsupervised methods is thus comparable to the supervised methods. This is surprising, as these unsupervised classifiers are not trained on any deepfake data --- they are trained on warped images that \textit{simulate} common deepfake artifacts. We hypothesize that this training strategy might have enabled better detection generalization capabilities.

\noindent \textbf{Detection performance on the \researchdataset{} dataset.} The fourth row in Table~\ref{tab:detection_performances} presents (best-attainable) detection performance for each classifier on \researchdataset{}. The performance of all classifiers is poor, and worse than their performance on \wilddataset{}.
We can attribute this failure to the same lack of generalization capabilities to which we attributed poor performance on the \wilddataset{} dataset.



The best-performing supervised classifier on \researchdataset{} is CapsuleForensics, with an F1 score of 66\%. Multi-Task exhibits a higher F1 score of 67\%, but this is due to the nature of the F1 metric, as MultiTask produces the same classification decision of `fake' for nearly all inputs. All supervised classifiers exhibit poor performances, with F1 scores as low as 55\%. The best-performing unsupervised classifier on \researchdataset{} is DSP-FWA.

We also present performance on each research dataset that constitutes \researchdataset{}, separately. These performances are presented in rows 5--10 of Table~\ref{tab:detection_performances}. No classifier performs consistently across each of the 6 individual sets. We note that CapsuleForensics and Xception perform exceptionally well on FFPP, as also reported in their original work.


\noindent \textbf{Deep diving into the performance of CapsuleForensics and DSP-FWA.}
We perform a deeper analysis of the best-performing supervised and unsupervised methods --- CapsuleForensics and DSP-FWA. First, we investigate the relationship between detection performance and the deepfake generation method,~\ie whether these classifiers are biased against certain deepfake generation methods. We thus measure their efficacy on individual subsets of \wilddataset{}, each corresponding to a different generation method. Table~\ref{tab:capsule_gen_methods_dist} shows the percentage of deepfake videos correctly classified (TP) and the percentage of videos wrongly classified as real (FN). We find that CapsuleForensics performs better against videos generated by DFL and Zao, compared to FakeApp and FaceSwap. It is also able to correctly classify 80\% of videos generated by `unknown' generation methods.
DSP-FWA performs significantly better on all but Zao deepfakes, with 10\% FN rate compared to 4\% or lower on all other categories. These results suggest that detection methods need to be tested against a variety of deepfake generation methods to understand their true performance. 



\begin{table}[h]
    \centering
    \begin{tabular}{c|c|c|c|c|c|c}
        \hline
        \bf Classifer & \bf Score & \bf Unknown & \bf DFL & \bf \makecell{Face \\ Swap} & \bf \makecell{Fake \\ App} & \bf Zao \\
        \hline
        \multirow{2}{*}{\makecell{Capsule \\ Forensics}} & FN & 20\% & 12\% & 36\% & 33\% & 12\%\\
        \cline{2-7}
        & TP & 80\% & 88\% & 64\% & 67\% & 88\% \\
            \hline
        \multirow{2}{*}{\makecell{DSP- \\FWA}} & FN & 4\% & 1\% & 0\% & 0\% & 10\%\\
        \cline{2-7}
        & TP & 96\% & 99\% & 100\% & 100\% & 90\% \\
        \hline
    \end{tabular}
    \vspace{1ex}
    \caption{Distribution of generation methods for videos that were correctly classified (TP) and incorrectly classified (FN) by CapsuleForensics and DSP-FWA.}    \label{tab:capsule_gen_methods_dist}
    \vspace{-0.1in}
\end{table}

\begin{figure}[t]
\centering
\begin{subfigure}[t]{0.48\columnwidth}
  \includegraphics[width=\columnwidth]{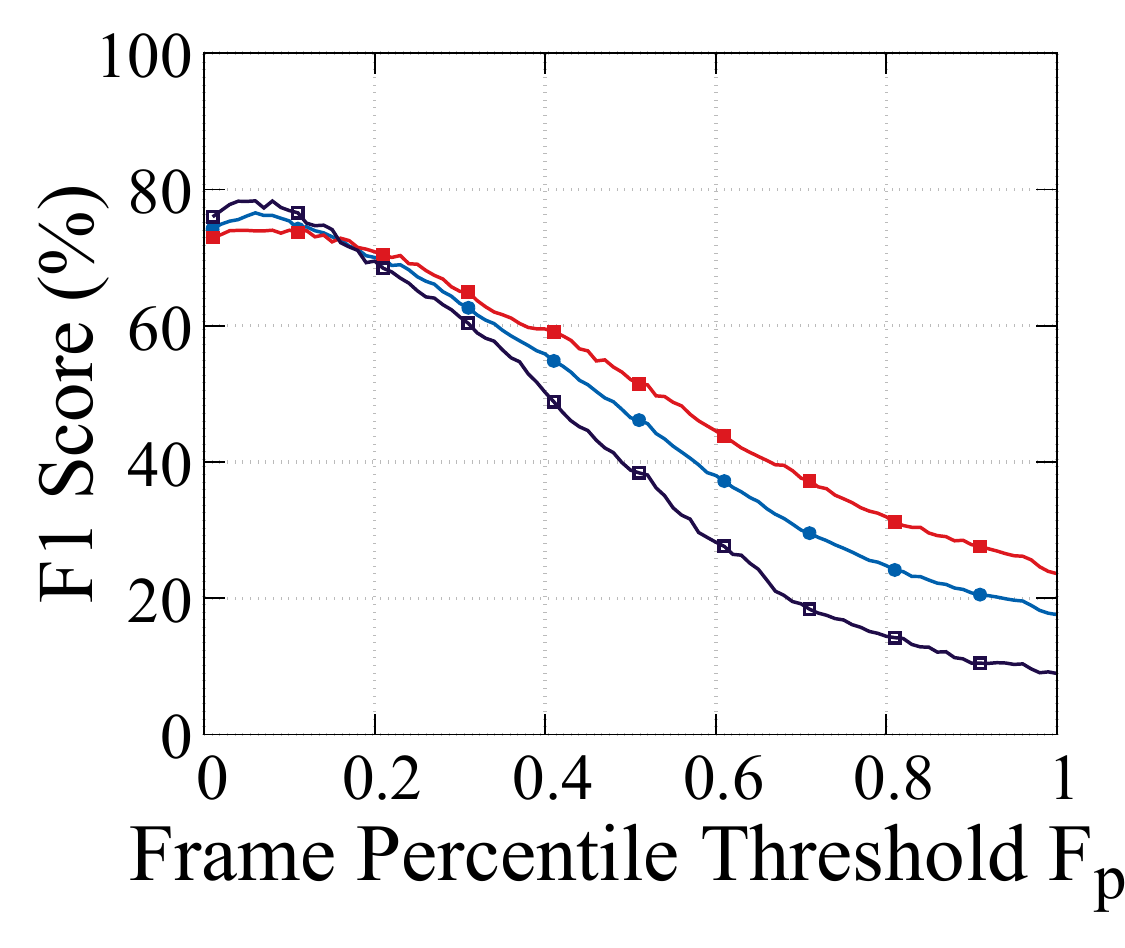}
\caption{}
\label{fig:capsule_dfw_f1}
\end{subfigure}
\hfill
\begin{subfigure}[t]{0.48\columnwidth}
  \includegraphics[width=\columnwidth]{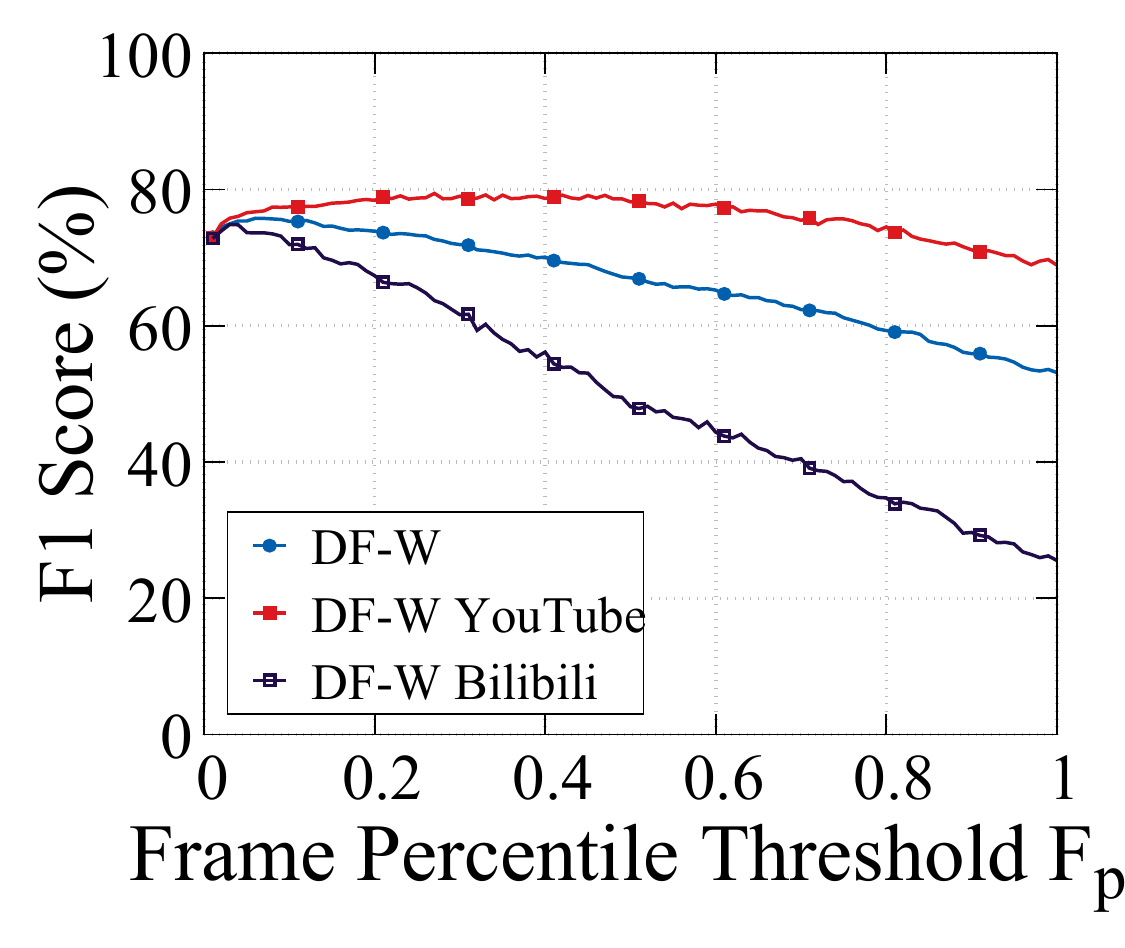}
\caption{}
\label{fig:dspfwa_dfw_f1}
\vspace{-2ex}
\end{subfigure}
\caption{\wilddataset{} F1 scores vs. $F_p$ for (a) CapsuleForensics (b) DSP-FWA.}
 \label{fig:f1_sweep}
\end{figure}

Next, we consider the role played by the choice of the $F_p$ decision threshold parameter. As discussed in Section~\ref{sec: experimental_setup}, this $F_p$ parameter is used for video-level decisions. 
Figures~\ref{fig:capsule_dfw_f1} and~\ref{fig:dspfwa_dfw_f1} shows the effect of the decision threshold $F_p$ on the F1 scores for CapsuleForensics and DSP-FWA, respectively, on the DF-W dataset. 
These sweeps reveal that for most cases, $F_p$ values roughly $<0.2$ are generally optimal. 

\noindent \textbf{Analyzing detection failures.} DNN-based facial recognition models are known to suffer from racial bias~\cite{robinson2020face}. Given that deepfake detection schemes are also based on DNN models that analyze faces, we investigate for any issues of racial bias. We focus on the best overall classifier, CapsuleForensics.


\noindent\textit{Racial bias.}
~We evaluate whether CapsuleForensics performs poorly for specific races. We prepare a test set of face images with race labels using the \wilddataset{} YouTube dataset, and containing a balanced number of fake and real images for each race.
To construct this test set, we extract the 10 most confidently predicted faces in each video, and assign it a race using DeepFace~\cite{serengil2020lightface} --- a facial analysis DNN that can predict the race for a given face image. We further discard faces for which DeepFace itself is below 95\% in  prediction confidence. This heuristic enables DeepFace to exhibit high precision (over 90\%), as verified on each race of a labelled racial classification dataset~\footnote{We use the FairFace dataset from Karkainnen et al.~\cite{karkkainen2021fairface}}. The resulting dataset comprises faces belonging to 3 races --- Caucasian, Black, and Asian. Finally, balancing is achieved by (randomly) downsampling the larger class (fake or real) for each of these 3 races.

The resulting race-wise detection performance of CapsuleForensics indicates a racial bias in its detection capabilities. \textit{We observe that the (macro average) F1 score for 464 Asian faces is only 48\%}. This performance is notably low when compared with performance for the other races: 72\% for 800 Caucasian faces, and 74\% for 416 Black faces.
To explain the cause of such a racial bias, a natural direction is to analyze the race distribution of \researchdataset{} FFPP,~\ie the CapsuleForensics training set. However, after adopting a precision optimized configuration for DeepFace, there is an inadequate number of face images remaining in \researchdataset{} FFPP to analyze. We thus leave  investigation into the cause of this racial bias as future work.

\para{Model interpretation schemes to understand detection decisions.}
Using CapsuleForensics as a case study, we focus on understanding the predictions made by the classifier using a model interpretation scheme. Besides the more immediate benefit of improving detection schemes, understanding relevant features in an image that led to a decision is also important in an adversarial context. Specifically, an adaptive adversary with knowledge of the relevant features for classification decisions can evade detection by eliminating (or spoofing) those relevant, critical features. 

To this end, we use Integrated Gradients~\cite{sundararajan2017axiomatic} or IntGrad, a recent and influential feature-attribution based DNN explanation methodology, to explain the outputs of CapsuleForensics. For a given model and an input image, IntGrad produces a saliency map representing an attribution score for every channel of every pixel in the input image. This map ``explains'' the output of the neural network. However, the maps themselves vary from image to image. To identify specific patterns and trends in these maps, we conduct a systematic, manual annotation process.


\noindent
\emph{Step 1: Sampling representative images for explanation.}
Representative real and deepfake faces are chosen from \wilddataset{} and \researchdataset{} respectively, and categorized into 4 classes --- \textit{TP}, \textit{TN}, \textit{FP}, \textit{FN} ---
based on the output of CapsuleForensics. For each video, one representative face is chosen. Specifically, for TP and FP videos, the face with maximum probability being fake is chosen, and for TN and FN videos, the face with maximum probability being real is chosen. In total, we sample 1,533, 1,100, 798, and 177 images from the TP, TN, FP, and FN classes, respectively.

\noindent
\emph{Step 2: Using IntGrad to obtain saliency masks.}
For each face image, we obtain feature attribution scores from IntGrad, considering the top 10\% pixels (according to IntGrad scores) to be the most important pixels, and produce saliency maps.

\begin{figure}[t]
    \centering
    \includegraphics[width=0.9\linewidth]{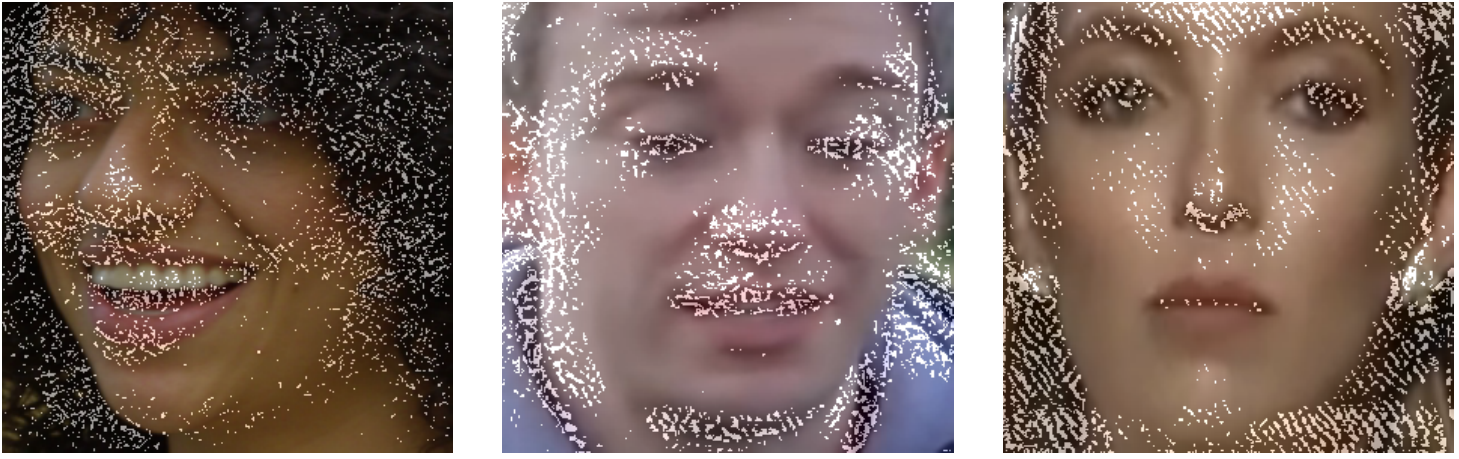}
    \caption{Visualization of salient regions (white dots) as determined by CapsuleForensics, and highlighted using IntGrad~\cite{sundararajan2017axiomatic}. The first image is a fake face correctly classified as fake, and shows no-discernible-features. The second image is a real face correctly classified as real, and shows face boundary features. The third image is a fake face classified as real, and shows salient background features. Note that these are \underline{significant} features, and do not preclude insignificant features such as those around the eyes and nose.}
    \label{fig:capsule_vis}
\end{figure}

\noindent
\emph{Step 3: Manually annotating the saliency maps.}
To identify pattern trends in the saliency maps, we conduct a manual annotation process. Categories are chosen to represent different parts of the face image being focused by the saliency masks. The initial categories chosen for this purpose include: eyes, nose, mouth, face-boundary, background, and no-discernible-features. The no-discernible-features category is special in that it is used to describe saliency masks in which no specific parts of the face were highlighted. To narrow down the set of categories, 30 face images from each of the four classes (TP, FP, TN, FN) are picked
and annotated using the above categories. A face image could be annotated with multiple categories. Each image is annotated independently by 3 of the co-authors of the paper. An image is finally annotated with a category, if at least 2 of the 3 annotators use the given category to annotate the saliency mask.
After this step, a difference in the total number of annotations between the TP, FP, TN, FN classes was observed for only 3 categories: no-discernible-features, background, and face-boundary. Figure~\ref{fig:capsule_vis} shows representative examples of faces annotated with the above 3 labels.~\footnote{Note that while other feature categories (eyes, mouth, etc) might be visible in these examples, the above process has eliminated them as relevant to the classifier's decision.} Next, to understand if the differences were statistically significant, we randomly sample 100 images per class and annotate them with the 3 categories. 


\emph{Statistical analysis of saliency mask annotations and implications.}
Based on the results of the annotation process, we find that compared to FN and TN, TP and FP classes have significantly higher number of faces annotated with no-discernible-features
(p=0.031, odds ratio=2.579)%
\footnote{All comparisons are done using Fisher's Exact test.
The p-values are reported after Bonferroni correction.},
and significantly lower number of faces annotated with background
(p=$2.89\cdot10^{-11}$, Odds Ratio=0.216).
This suggests that if an adversary manages to introduce relevant background features just around a deepfake face,
CapsuleForensics would be $\approx 4.6$ times as likely to label the image as real.
From the opposite perspective, it is to the benefit of the defender to avoid focusing outside of the facial boundary,
and ensure that the feature extraction module produces relevant features.
From this defender's perspective, it is also important to note that when CapsuleForensics is unable to focus on important features (no-discernable-feature case)
it is $\approx 2.5$ times more likely to label the image as fake, independent of its true source.
Ensuring the model is able to focus on proper features might be achieved
by pre-emptively validating/improving the feature extraction scheme.
Finally, compared to the other three classes, TN class has significantly higher number of faces annotated with face-boundary
(p=0.002, odds ratio=2.627).
This seems to further support the idea that
being able to correctly identity facial features and face boundary
is critical for the classifier in being able to discern real faces from the fakes.

\subsection{Towards Improving Detection Performance}
In light of the poor generalization capabilities of existing detection schemes, we consider multiple approaches for improving performance, and evaluate their efficacy. Table~\ref{tab:detection_performances_improved} presents the results of this evaluation.

\begin{table*}[h]
    \centering
    \begin{tabular}{l|>{\columncolor[rgb]{0.83,0.83,0.83}}c|c|c|>{\columncolor[rgb]{0.83,0.83,0.83}}c|c|c|>{\columncolor[rgb]{0.83,0.83,0.83}}c|c|c|>{\columncolor[rgb]{0.83,0.83,0.83}}c|c|c|>{\columncolor[rgb]{0.83,0.83,0.83}}c|c|c}
    \hline
    \multirow{3}{*}{\bf Dataset} 
    & \multicolumn{3}{c|}{\bf Capsule - DA} & \multicolumn{3}{c|}{\bf Capsule - SE} & 
    \multicolumn{3}{c|}{\bf MesoNet - DA} & \multicolumn{3}{c|}{\bf MesoNet - SE} &  \multicolumn{3}{c}{\bf Seferbekov} \\
    \cline{2-16}
    & \bf F1 & \bf P & \bf R & \bf F1 & \bf P & \bf R & \bf F1 & \bf P & \bf R & \bf F1 & \bf P & \bf R & \bf F1 & \bf P & \bf R \\
    \hline
    \wilddataset{} & \underline{\bf91} & \bf 87 & \bf 94 & \bf 71 & \bf 60 & \bf 87 & \bf 62 & \bf 48 &\bf 88& \bf 67& \bf 50& \bf 100 & \bf81 & \bf 71 & \bf 93\\
        \wilddataset{} YouTube & \underline{90} & 88 & 92 & 72 & 64 & 83 &  62 &48 &87 & 66& 50& 100& 85 & 82 & 89 \\
    \wilddataset{} Bilibili &  \underline{94} & 89 & 98  &71 & 59 & 89 &  63 &49 & 89&67& 51& 100& 79 & 69 & 94\\
    \hline
    \researchdataset{} & \bf 65 & \bf 58 &\bf 74& \underline{\bf93} & \bf 91& \bf 94& \bf 61& \bf 48& \bf 83 &\bf 62&\bf 48&\bf 89& \bf89  & \bf 86 &\bf 91 \\
    \researchdataset{} FFPP & 91 & 94 & 88 & \underline{95} &93&98 &38 &  33& 46&46&38&60& 91 & 87 & 97\\
    \researchdataset{} Celeb-DF & 75 & 67 & 84 &\underline{91} & 90 & 93 & 65 & 50 &93 &66&49& 98& \underline{91}  & 88 & 94\\
    \researchdataset{} DFDC & 48 & 46 & 50 &\underline{93}&90& 95& 65& 50& 93&64&49& 95& 90 & 88 & 92 \\
    \researchdataset{} DFD & 81 & 79 & 83 &\underline{95}&92&98& 63 & 49& 90 &65&49&96&  92 & 87 &97 \\
    \researchdataset{} DTimit & 14 & 17 &12 &\underline{92}&94&91&72 & 59& 92 &66&50&99& 79  &84  &74 \\
    \researchdataset{} UADFV & 90 & 95 & 86 &\underline{96} & 93 & 98 &56 & 45 &76  &67&51&100& 95 & 96 & 94\\
    \hline
    \end{tabular}
    \vspace{1ex}
    \caption{Best attainable detection performances (F1) of Capsule-DA, Capsule-SE, MesoNet-DA, MesoNet-SE and DFDC winner Seferbekov's model on the \wilddataset{} and \researchdataset{} Datasets, with corresponding precision (P) and recall (R) scores. Here `DA' refers to models after domain adaptation retraining, and `SE' refers to models after source expansion retraining. Retraining is achieved by fine-tuning the whole model, \ie without freezing layers. Best attainable detection performance (F1) for each dataset is underlined.} 
    \label{tab:detection_performances_improved}
    \vspace{-4ex}
\end{table*}

\noindent \textbf{Improving detection results via transfer learning.} So far, we directly used the pre-trained model provided by the authors of each detection scheme. This helps to understand the practical applicability of these pre-trained models, which unfortunately indicated poor performance out-of-the-box. To improve performance, we investigate the use of \textit{transfer learning}. Starting from the pre-trained weights, we retrain a given model using a limited amount of additional new data. 





Specifically, we study two retraining strategies --- \textit{(1) Domain adaptation (DA) retraining (2) Source expansion (SE) retraining.} For DA retraining, we assume access to a small set of videos from DF-W for transfer learning, thus building the potential capability to adapt to the new domain of in-the-wild deepfake videos~\cite{kouw2018introduction}. We only assume access to a limited number of deepfake videos (around 50), because in practice, it will be hard to obtain access to a large number of deepfake videos created for misuse. In SA retraining, we include new videos from multiple academic deepfake datasets to expand the training data distribution. Overall, our goal is to increase the generalizability of the detection model by learning a new or a boarder data distribution provided by the retraining dataset.


\noindent \textit{Retraining configuration.} We choose the two best supervised detection methods, \ie CapsuleForensics and MesoNet for our analysis. Retraining the models involves initialization with the pre-trained weights, followed by fine-tuning of all model layers until convergence.
For domain adaptation, the retraining dataset contains 50 deepfake videos randomly sampled from \wilddataset{} (40 YouTube and 10 Bilibili), and 120 real videos evenly and randomly sampled from each of the 6 academic research subsets in \researchdataset{}. 
To extract fake faces (since there are unaltered faces in \wilddataset{} videos), we manually scan through videos to collect clips with only synthetic faces that add up to 10 seconds from each deepfake video. From collected clips of each video, 100 faces are randomly sampled. To extract real faces, we randomly sample 50 faces from each real video.
Finally, we obtain a retraining dataset with 6,000 real faces and 5,000 fake faces. 
For source expansion, the retraining dataset contains 240 deepfake video and 240 real videos evenly and randomly sampled from each of the 6 academic research subsets in \researchdataset{}. We randomly sample 40 faces from each video, and obtain a retraining dataset with 9,600 real and 9,600 fake faces.   

\noindent \textit{Results.} Detection performances post-retraining are presented in Table~\ref{tab:detection_performances_improved}. We make the following key observations: \textit{First}, CapsuleForensics-DA (domain adaptation) improves performance on \wilddataset{} to 91\% F1 score (from 77\% F1). \textit{This indicates that domain adaptation using transfer learning is a promising approach towards improving performance, while requiring only a small number of deepfake videos from the target domain.}
However, we observe some fluctuation in performance for CapsuleForensics-DA when applied to the DF-R datasets. For 3 out of the 6 DF-R datasets, there is a drop in performance. To limit such performance degradation, we tried reducing the number of layers fine-tuned during transfer learning. We observe that performance degradation on DF-R datasets can be limited, if we only fine-tune the top 30 layers (out of 160 layers), instead of fine-tuning all layers. For example, performance on DF-R DTimit bounces back from 14\% to 52\%. However, there is a small performance hit on DF-W on selectively fine-tuning layers---F1 score on DF-W reduces from 91\% to 87\%.

\textit{Second}, for DA, CapsuleForensics performs better than MesoNet, and performance of MesoNet drops compared to the setting without domain adaptation. We suspect that this is because MesoNet has less information capacity than CapsuleForensics due to its smaller architecture, and is not well suited for transfer learning in this setting. The Meso4 network only has 27,977 trainable parameters~\cite{nguyen2019use}, while the CapsuleForensics network has 3,896,638 trainable parameters~\cite{afchar2018mesonet}. This suggests that not all methods can be improved using transfer learning.

\textit{Third}, while the domain adaptation strategy works well, the source expansion (SE) strategy does not improve performance on DF-W datasets. Therefore, simply combining existing academic deepfake datasets cannot capture the distribution of \wilddataset{}. However, CapsuleForensics-SE (source expansion) effectively improves performance on \researchdataset{} to 93\% F1 score (>90\% on all subsets), which is expected as we have retrained the model to better adapt to the DF-R datasets.



\para{How well does the best model from a public deepfake detection competition work?} 
Another approach to building better detection schemes
is to outsource the task to the public, through funded competitions. We examine the practical applicability of a competition winning model,  by leveraging Facebook's recently concluded DFDC competition~\cite{dfdc_competition}.
The competition concluded in July 2020, with 5 detection schemes selected as prize-winners out of more than 2000 submissions. We choose the best model, henceforth referred to as the Seferbekov model~\cite{selim_github} (named after the author), and evaluate it on all datasets.

The Seferbekov model was trained on the public training set of DFDC dataset~\cite{dolhansky2020deepfake} released by Facebook, which includes 100,000 deepfake clips and 19,154 real clips. This model uses the state-of-the-art EfficientNet B7~\cite{tan2019efficientnet} (pretrained with ImageNet and noisy student) for feature encoding. Structured parts of faces were dropped during training as a form of augmentation.
For our evaluation, we use the pre-trained model provided by the author.~\footnote{\url{https://github.com/selimsef/dfdc_deepfake_challenge/blob/master/download_weights.sh}} In the competition, the Seferbekov model makes use of predictions based on 32 frames from each video to make a video-level prediction.
However, 
this strategy can potentially deteriorate the detection performance on \wilddataset{} as \wilddataset{} videos contain unaltered frames and have longer duration compared to deepfake clips tested in the competition. Thus we adopt the strategy proposed in Section~\ref{sec: experimental_setup} to make the video-level decision. 

The evaluation results are presented in Table~\ref{tab:detection_performances_improved}. We make the following observations: \textit{First}, the winning model (out of 2000 submissions) still does not exhibit very high detection performance (\eg F1 score > 90\%) on the DF-W dataset. The Seferbekov model achieves an 81\% F1 score, but with a low precision of 71\%. This suggests that there is still room for significant improvement, before we are ready for real-world deployment. \textit{Second}, domain adaptation and source expansion techniques outperform the winning model on DF-W, and DF-R datasets, respectively.
\textit{Third}, the Seferbekov model boasts relatively better performance (F1) on both \wilddataset{} and \researchdataset{} when compared to the existing classifiers (without DA or SE), shown in Table~\ref{tab:detection_performances}. There are three potential reasons for this: (1) the Seferbekov model was trained on the large, state-of-the-art DFDC public training dataset~\cite{dolhansky2020deepfake} comprising 100,000 deepfake videos, whereas CapsuleForensics was trained on the smaller FaceForensics++ dataset comprising only 7,480 deepfake and other types of manipulated videos. (2) According to its official GitHub repository~\cite{selim_github}, the Seferbekov model applied heavy data augmentation on the provided DFDC training set, including out-of-the-box image augmentations in Albumentations~\footnote{https://github.com/albumentations-team/albumentations} library and face part cut-out. (3) the Seferbekov model averages predictions from 7 models trained with different seeds, whereas CapsuleForensics uses a single model.

\section{Limitations of our study}
\label{sec:limitations}

Our study has a few limitations. (1) The proposed \wilddataset{} dataset should not be mistaken for one that is representative of all deepfake videos on the Internet.  It is very likely that other deepfake generation methods exist, for which we were unable to find videos in the wild. For example, there are deepfake methods that can synthesize an entire face using GAN-based approaches~\cite{karras2019style}.  (2) Our study did not investigate pornographic deepfakes in the wild, and leave their investigation as future work. Many such `not safe for work' deepfake videos exist on the Internet. 
However, it is worth noting that the prolific deepfake pornography forum \textit{MrDeepFakes~\footnote{\url{https://mrdeepfakes.com/forums/}}} describes a near universal adoption of DFL as their choice of deepfake generation method. Videos generated using this method constitute over 20\% of \wilddataset{}. We thus hope that insights gained from \wilddataset{} might partially transfer to more illicit domains. (3) There are other deepfake detection methods for which we were unable to obtain pre-trained models. These methods include Two-stream~\cite{zhou2017two}, HeadPose~\cite{yang2019exposing}, which are supervised models, and 
FaceXray~\cite{li2019face}, Camera-based Fingerprints~\cite{cozzolino2019extracting} that are unsupervised models.
(4) Deepface, the race prediction DNN used in analyzing detection failures, might not perform optimally. We attempt to minimize the effects of any such sub-optimal racial classification by using a precision optimized version of the classifier (see Section~\ref{sec:detection-performance}). 

\section{Conclusion}
We presented a measurement and analysis study of deepfake videos found in the wild. We collected and curated a novel dataset, \wilddataset{}, comprising 1,869 deepfake videos ---
the largest dataset of real-world deepfake videos to date. 
Our analysis revealed that \wilddataset{} videos differ from the deepfake videos in existing research community datasets in terms of content, and generation methods used, raising new challenges for detection of deepfake videos in the wild. 
We further systematically evaluated multiple state-of-the-art deepfake detection schemes on \wilddataset{},
revealing poor detection performance. This suggests a distributional difference between in-the-wild deepfakes, and deepfakes in research community datasets.
We also attributed detection failures to be related to racial biases, and using model interpretation schemes, we investigated features that can be leveraged to either improve or evade detection.
Finally, we show that domain adaptation via transfer learning, which retrains the model on a small set (50 videos) of \wilddataset{} videos, is a promising approach to improving performance on \wilddataset{}. Overall, our findings indicate a need for incorporating in-the-wild deepfake videos into future work.


\bibliographystyle{ACM-Reference-Format}
\bibliography{reference}

\appendix 
\section{Ethics}
\noindent We did not create any deepfake videos ourselves, or further share existing deepfake videos on the Web. We did not create or share any deepfake generation tools on the Web or with any entity or user. All data and videos we downloaded were publicly available information. We did not collect or study ``not safe for work'' deepfake videos (\eg pornographic or explicit content). 

\end{document}